\documentclass{ws-mplb}
\usepackage{amsfonts}
\usepackage{amsmath}
\usepackage{amssymb}
\usepackage{graphicx}

\begin{document}

\markboth{Velazquez and Curilef}{Uncertainty relations of
Statistical Mechanics}

%
\catchline{}{}{}{}{}
%

\title{UNCERTAINTY RELATIONS OF STATISTICAL MECHANICS}

\author{\footnotesize LUISBERIS VELAZQUEZ\footnote{\textit{Present address}: Departamento de F\'{\i}sica, Universidad Cat\'{o}lica del Norte, Av. Angamos
0610, Antofagasta, Chile.}}

\address{Departamento de F\'{\i}sica, Universidad de Pinar del R\'{\i}o \\
 Mart\'{\i} 270, Esq. 27 de Noviembre, Pinar del R\'{\i}o, Cuba.\\
lvelazquez@ucn.cl}

\author{SERGIO CURILEF}

\address{Departamento de F\'{\i}sica, Universidad Cat\'{o}lica del Norte\\ Av. Angamos
0610, Antofagasta, Chile.\\
scurilef@ucn.cl}

\maketitle

\begin{history}
\received{(Day Month Year)}
\revised{(Day Month Year)}
\end{history}

\begin{abstract}
Recently, we have presented some simple arguments supporting the
existence of certain complementarity between thermodynamic
quantities of temperature and energy, an idea suggested by Bohr and
Heinsenberg in the early days of Quantum Mechanics. Such a
complementarity is expressed as the impossibility of perform an
exact simultaneous determination of the system energy and
temperature by using an experimental procedure based on the thermal
equilibrium with other system regarded as a measure apparatus
(thermometer). In this work, we provide a simple generalization of
this latter approach with the consideration of a thermodynamic
situation with several control parameters.
\end{abstract}


\section{Introduction}

Bohr and Heisenberg suggested in the past that the thermodynamical
quantities of temperature and energy are complementary in the same
way as position and momentum in Quantum
Mechanics\cite{bohr,Heisenberg}. Their argument was that a definite
temperature can be attributed to a system only if it is submerged in
a heat bath, in which case energy fluctuations are unavoidable. On
the other hand, a definite energy can be assigned only to systems in
thermal isolation, thus excluding the simultaneous determination of
its temperature. By considering the analogy with Quantum Mechanics,
a simple dimensional analysis allows to conjecture the following
uncertainty
relation:%
\begin{equation}
\Delta U\Delta (1/T)\geq k_{B},  \label{pur}
\end{equation}%
where $k_{B}$ is Boltzmann' s constant.

A serious attempt in order to support the expression (\ref{pur}) was
provided by Rosenfeld\cite{Rosenfeld} in turn of 1960' within the
framework of classical fluctuation theory\cite{Landau}. However,
this approach was performed under special restrictions which meant
that the fluctuations of energy and temperature became dependent on
each other and were no longer really complementary. Along the years,
other formulations of the thermodynamical uncertainty relations were
proposed by Mandelbrot\cite{Mandelbrot}, Gilmore\cite{Gilmore},
Lindhard\cite{Lindhard}, Lavenda\cite{Lavenda},
Sch\"{o}lg\cite{Scholg}, among other authors. Remarkably, the
versions of this relation which have appeared in the literature give
different interpretations of the uncertainty in temperature $\Delta
\left( 1/T\right) $ and often employ widely different theoretical
frameworks, ranging from statistical thermodynamics to modern
theories of statistical inference. Despite of all devoted effort,
this work has not led to a consensus in the literature, as clearly
discussed in the most recent review by J. Uffink and J. van
Lith\cite{Uffink}. Even nowadays, this intriguing problem is still
lively debated\cite{Lavenda2,Uffink2}.

In our opinion, the underlying difficulties in arriving at a
definitive formulation of the energy-temperature complementarity
rely on a common and subtle misunderstanding of the
\textit{temperature concept}. Most of previous attempts have tried
to justify a complementarity between energetic isolation and thermal
contact inspired on the Bohr's intuitive arguments. A primary idea
here is that a system has a \textit{definite temperature} when it is
put in thermal contact with a heat bath (a system having an infinite
heat capacity) at that temperature, which is the equilibrium
situation associated to the known Gibbs' canonical ensemble:
\begin{equation}
p_{c}\left( U\left\vert \beta \right. \right) dU=Z\left( \beta
\right) ^{-1}\exp \left( -\frac{1}{k_{B}}\beta U\right) \Omega
\left( U\right) dU,\label{can}
\end{equation}%
where $\beta =1/T$ , $\Omega \left( U\right) $, the states density,
and $Z\left( \beta \right) $, the partition function. If one can
only attribute a definite value of the system temperature by
appealing to the temperature of a second thermodynamic system (the
heat bath), this fact necessarily implies that the temperature of an
isolated system is a meaningless concept, or at least, it is
imperfectly definite. Such an idea is explicitly expressed in some
classical books on Statistical Mechanics, as the known Lev Landau
and Evgenii Lifshitz treatise\cite{Landau} (see last paragraph of \S
112). This latter conclusion is counterfactual, since it could not
be possible to talk about a definite value for the temperature of
the system acting as a heat bath when it is put in energetic
isolation. Besides, the actual meaning of the \textit{temperature
uncertainty} $\Delta \left( 1/T\right) $ in Eq.(\ref{pur}) is
unclear from this viewpoint, since the temperature $T$ is just a
constant parameter of the canonical probabilistic distribution
(\ref{can}). This particular question leads to the polemic exchange
among Feshbach\cite{Feshbach}, Kittel\cite{Kittel} and
Mandelbrot\cite{MandelT} in the years 1987-1989 in Physics Today,
and explains by itself the several interpretations of temperature
uncertainty $\Delta (1/T)$ existing in the literature.

On the contrary, the concept of temperature of a closed isolated
system admits a clear and unambiguous definition in terms of the
microcanonical temperature derived from the celebrated Boltzmann's
entropy:
\begin{equation}\label{BE}
S=k_{B}\log W\rightarrow\frac{1}{T}=\frac{\partial S}{\partial U}.
\end{equation}
One can realize after revising the Gibbs' derivation of canonical
ensemble (\ref{can}) from the microcanonical ensemble that the
temperature appearing as a parameter in the canonical distribution
(\ref{can}) is just the microcanonical temperature of the heat bath
when its size is sent to the thermodynamic limit
$N\rightarrow\infty$. Consequently, such a parameter actually
characterizes the internal thermodynamical state of the bath and its
thermodynamic influence on the system under consideration. While the
differences between the temperature $T$ appearing in the Gibbs's
canonical distribution (\ref{can}) and the one associated to the
microcanonical ensemble of a closed system (\ref{BE}) is not so
relevant in most of practical situations involving large
thermodynamic systems driven by short-range forces, this is not the
case of small or mesoscopic systems such as molecular and nuclear
clusters, or even, the large thermodynamic system driven by
long-range interactions such as astrophysical
systems. The crucial feature is that only by using the microcanonical temperature (%
\ref{BE}) it is possible to describe the existence of
\textit{negative heat capacities} $C<0$ and the occurrence of
\textit{phase transitions in finite
systems}\cite{moretto,Dagostino,gro na,gro1,Dauxois}.

By using this latter interpretation, temperature is, at least in a
formal viewpoint, a function on the energy of a closed system.
Consequently, these quantities seem to be no longer complementary.
Indeed, the imposition of the energetic isolation $\Delta U=0$ into
Eq.(\ref{BE}) leads to $\Delta \left(
1/T\right) =0$, which clearly violates the supposed validity of Eq.(\ref{pur}%
). Apparently, there is not way to support the existence of a
complementary relation between the energy and temperature by
assuming microcanonical definition (\ref{BE}). Fortunately, this
preliminary conclusion is incorrect. The temperature is not a
physical quantity with a direct \textit{mechanical interpretation}
as the energy. As the entropy, it is just a
\textit{thermo-statistical} quantity whose physical interpretation
is only possible by appealing to the notion of \textit{statistical
ensemble}. Such a nature implies that any practical determination of
the temperature of a system is always imprecise. In practice, a
temperature mensuration can be only performed, in an indirectly way,
through the statistical processing (or the temporal expectation
values) of certain physical observables commonly referred as
\textit{thermometric quantities}. Besides, it is necessary to
account for the \textit{unavoidable affectation} produced by the
experimental measurements on the internal state of the system under
study, a feature which is simply ignored in previous attempts to
justify an energy-temperature complementarity.

Recently, we reconsider this old problem of Statistical Mechanics by
analyzing the limits of precision involved during the practical
determination of the temperature of a given system by using the
usual experimental procedure based on the thermal equilibrium with a
second system\cite{VelazquezC,Vel}, which plays the role of a
measure apparatus (thermometer). Our analysis allows to obtain the
following result:
\begin{equation}
\Delta U\Delta \eta \geq k_{B},  \label{tur}
\end{equation}%
where $\eta =1/T_{A}-1/T $ the inverse temperature difference
between the measure apparatus and the system under study, and
$\Delta a=\sqrt{\left\langle \delta a^{2}\right\rangle }$, the
square root of the statistical deviation of a physical quantity $a$
undergoing thermal fluctuations.

A crucial difference of result (\ref{tur}) with other attempts to
justify the complementary character between energy and temperatures,
as the cases of Rosenfeld's and Mandelbrot's
approaches\cite{Rosenfeld,Mandelbrot}, is found in the fact that the
thermal uncertainties $\Delta U$ and $\Delta \eta $ depend on the
nature of the measure apparatus. Thus, an experimentalist has a free
will to change the experimental conditions and modify the energy and
temperature uncertainties. Clearly, Eq.(\ref{tur}) indicates the
impossibility of carry out an exact simultaneous determination of
the energy and temperature of a given system by using any
experimental procedure based on the consideration of the thermal
equilibrium condition with a second thermodynamic system.

As already evidenced, our proposal of energy-temperature
complementarity is remarkably simple, and even, it is quite an
expected result. Clearly, the physical arguments leading to the
uncertainty relation (\ref{tur}) admit a simple extension in order
to support a complementary character among other conjugated
thermodynamic variables. The analysis of this question is the main
goal of this work.

\section{Extending the energy-temperature complementarity}

\subsection{Notation conventions}

Conventionally, an equilibrium situation with several thermodynamic
variables is customarily described within the Statistical Mechanics
in terms
of the Boltzmann-Gibbs distributions\cite{Landau}:%
\begin{equation}
dp_{BG}\left( \left. U,X\right\vert \beta ,Y\right) =Z^{-1}\exp \left[ -\frac{%
1}{k_{B}}\beta \left( U+YX\right) \right] \Omega dUdX,  \label{BG}
\end{equation}%
being $\beta =1/T$, $Z=Z\left( \beta ,Y\right) $ the partition function and $%
\Omega =\Omega \left( U,X\right) $ the density of states. The quantities $%
X=\left( V,M,P,N_{i},\ldots \right) $ represent other macroscopic
observables acting in a given application like the volume $V$, the
magnetization $M$ and polarization $P$, the number of chemical species $%
N_{i} $, etc.; being $Y=\left( p,-H,-E,-\mu _{i},\ldots \right) $
the
corresponding conjugated thermodynamic parameters: the external pressure $p$%
, magnetic $H$ and electric $E$ fields, the chemical potentials $\mu
_{i}$, etc.

In order to simplify the analysis, we shall adopt the following
notation:
\begin{equation}
\left( U,X\right) \rightarrow I=\left( I^{1},I^{2},\ldots
I^{n}\right)
\end{equation}%
for the physical observables and
\begin{equation}
\left( \beta ,\xi =\beta Y\right) \rightarrow \beta =\left( \beta
_{1},\beta _{2},\ldots \beta _{n}\right)
\end{equation}%
for the thermodynamic parameters. Hereafter, we also assume the
Einstein's
summation convention:%
\begin{equation}
a^{i}b_{i}\equiv \sum_{i=1}^{n}a^{i}b_{i}.
\end{equation}%
The above notations allow to rephrase the distribution function (\ref{BG}%
) as follows:%
\begin{equation}
dp_{BG}\left( \left. I\right\vert \beta \right) =Z\left( \beta
\right) ^{-1}\exp \left[ -\frac{1}{k_{B}}\beta _{i}I^{i}\right]
\Omega \left( I\right) dI.
\end{equation}

\subsection{Starting considerations}

As elsewhere discussed, the Boltzmann-Gibbs distribution (\ref{BG})
accounts for an equilibrium situation between the system under study
is under the influence of a very large surrounding (a heat bath, a
particles reservoir, etc.). The thermodynamic variables $\left(
\beta ,Y\right) $ characterizing the internal thermodynamic states
of such a surrounding are regarded as constant parameters in
distribution (\ref{BG}). Such an approximation follows from the fact
that the size of the surrounding is so large that it is possible to
dismiss the thermodynamic influence of the system under study.

In general, the above equilibrium situation is unappropriated in
order to describe a general equilibrium situation between a system
(S) and certain \textit{measure apparatus} (A). Such an apparatus
constitutes a key piece in any experimental setup used to obtain the
thermodynamic parameters of the
system through the known \textit{equilibrium conditions}:%
\begin{equation}
\beta ^{S}=\beta ^{A},~Y^{S}=Y^{A}  \label{eq.cond}
\end{equation}%
(thermal equilibrium, mechanical equilibrium, chemical equilibrium,
etc.). Actually, the size of the interacting part of the apparatus
measure should be comparable or smaller than the system under study
in order to guarantee that the unavoidable interaction involved in
any measurement process does not affect in a significant way the
internal thermodynamic state of the system, e.g.: a thermometer
always interchanges energy with the system during the thermal
equilibration, and hence, it is desirable that its size be small in
order to reduce the perturbation of the system temperature.

Let us consider an equilibrium situation where the system and the
measure apparatus constitutes a closed system. By admitting only
additive physical observables (e.g.: energy, volume, particles
number or electric charge) obeying the constrain $I_{T}=I+I_{A}$, a
simple ansatz for the distribution function is given by:

\begin{equation}
p\left( \left. I\right\vert I_{T}\right) dI=\frac{1}{W\left( I_{T}\right) }%
\Omega _{A}\left( I_{T}-I\right) \Omega \left( I\right) dI,
\label{p1}
\end{equation}%
where $\Omega \left( I\right) $ and $\Omega _{A}\left( I_{A}\right)
$ are the densities of states of the system and the apparatus
measure respectively, and $W\left( I_{T}\right) $ the partition
function that
ensures the normalization condition:%
\begin{equation}
\int_{\Gamma }p\left( \left. I\right\vert I_{T}\right) dI=1,
\label{norm}
\end{equation}%
Here, $\Gamma $ the subset of all admissible values of the physical
observables $I$, which constitutes a compact subset of the Euclidean
n-dimensional space $R^{n}$, $\Gamma \subset R^{n}$, where $n$ is
the number
of macroscopic observables\footnote{%
We are dismissing in this approach the possible discrete nature of
the macroscopic observables $I$ associated to quantum effects, as
example, the quantification of the system energy, which is a
suitable approximation in
Statistical Mechanics when the systems under consideration are large enough.}%
.

The ansatz (\ref{p1}), however, clearly dismisses some important
practical situations where the additive constrain $I_{T}=I+I_{A}$
cannot be ensured for all physical observables involved in the
system-apparatus thermodynamic interaction. Significant examples of
\textit{unconstrained observables} are the magnetization or the
electric polarization associated to magnetic and electric systems
respectively. A formal way to overcome such a difficulty is carried
out by representing the density of state of the measure apparatus
$\Omega _{A}$ in terms of the physical observables of the system $I$
under study and certain set of control parameters $a$ determining
the given experimental setup, $\Omega _{A}=\Omega _{A}\left( \left.
I\right\vert a\right) $. Such an explicit dependence of the density
of states of the measure apparatus on the internal state of the
system follows from their mutual interaction, which leads to the
existence of \textit{correlative effects} between these systems.
Thus, our analysis starts from the consideration of the following
distribution function:
\begin{equation}
p\left( \left. I\right\vert a\right) dI=\frac{1}{W\left( a\right)
}\Omega _{A}\left( \left. I\right\vert a\right) \Omega \left(
I\right) dI. \label{ansatz}
\end{equation}%
Obviously, the subset $\Gamma $ of all admissible values of the
physical observables $I$ shall also depend on the control parameter
$a$, so that, it
is more appropriate to denote this subset as $\Gamma _{a}$ instead of $%
\Gamma $.

Let us denote by $p_{a}\left( I\right) \equiv p\left( \left.
I\right\vert a\right) $ the distribution function (\ref{ansatz})\
and $\partial _{i}A\left( I\right) =\partial A\left( I\right)
/\partial I^{i}$ the first
partial derivatives of a given real function $A\left( I\right) $ defined on $%
\Gamma _{a}$, where the positive integer $i\in \left[ 1,2,\ldots ,n\right] $%
. The general mathematical properties of the distribution function $%
p_{a}:\Gamma _{a}\rightarrow R$ associated to its thermo-statistical
relevance are the following:

\begin{enumerate}
\item[C1.] \textit{Existence}: The distribution function $p_{a}$ is a
nonnegative, bounded, continuous and differentiable function on $\Gamma _{a}$%
.

\item[C2.] \textit{Normalization}: The distribution function $p_{a}$ obeys
the normalization condition:%
\begin{equation}
\int_{\Gamma _{\alpha }}p_{a}\left( I\right) dI=1.
\end{equation}

\item[C3.] \textit{Boundary conditions}: The distribution function $p_{a}$
vanishes with its first partial derivatives $\left\{ \partial
_{i}p_{a}\right\} $ on the boundary $\partial \Gamma _{a}$ of the subset $%
\Gamma _{a}$. Moreover, the distribution function $p_{a}$
satisfies the condition:%
\begin{equation}
\lim_{\left\vert I\right\vert \rightarrow \infty }\left\vert
I\right\vert ^{\alpha }p_{a}\left( I\right) =0\text{ with }\alpha
\leq n+1 \label{cond.bound}
\end{equation}%
whenever the boundary $\partial \Gamma _{a}$ contains the infinite point $%
\left\{ \infty \right\} $ of $R^{n}$.

\end{enumerate}

The conditions (C1) and (C2) are natural for any distribution
function. The validity of condition (C3) accounts for the fact that
the product of densities of states $\Omega _{A}\left( \left.
I\right\vert a\right) \Omega
\left( I\right) $ usually vanishes at a finite point on the boundary $%
\partial \Gamma _{a}$ of the set $\Gamma _{a}$. The applicability of the
associated condition shown in Eq.(\ref{cond.bound}) ensures the
existence of the averages $\left\langle I^{i}\right\rangle $ and
some correlations functions when the boundary $\partial \Gamma _{a}$
contains the infinite point $\left\{ \infty \right\} $ of $R^{n}$.

\subsection{Thermodynamic parameters}

As elsewhere discussed in any standard book of Statistical Mechanics\cite%
{Landau}, thermodynamic parameters can be derived from the known
Boltzmann's entropy:

\begin{equation}
S=k_{B}\log W,
\end{equation}%
in terms of its first derivatives:%
\begin{equation}
\beta =\frac{\partial S\left( U,X\right) }{\partial U},~\beta Y=\frac{%
\partial S\left( U,X\right) }{\partial X},  \label{micro}
\end{equation}%
being $W=\Omega \delta c$ the coarsed grained volume, and $\delta c$
certain small energy constant. Such parameters provide the
thermo-statistical interpretation of the called equilibrium
conditions between two interacting separable systems,
Eq.(\ref{eq.cond}). Ordinarily, these relations follow from the
stationary conditions associated to the \textit{most likely
macrostate}.

The application of this latter argument for the case of distribution
function (\ref{ansatz}) yields:

\begin{equation}
k_{B}\partial _{i}\log p_{\alpha }\equiv \partial _{i}S_{A}+\partial
_{i}S_{S}=0,  \label{preli}
\end{equation}%
which can be rephrased as follows:

\begin{equation}
\beta _{i}^{S}=\beta _{i}^{A},  \label{cond.eq}
\end{equation}%
after assuming that the conjugated thermodynamic parameters of the system $%
\left\{ \beta _{i}^{S}\right\} $ and the measure apparatus $\left\{
\beta _{i}^{A}\right\} $ for both constrained as well as
unconstrained observables are defined by:
\begin{equation}
~\beta _{i}^{S}=\frac{\partial S_{S}\left( I\right) }{\partial
I^{i}},~\beta _{i}^{A}=-\frac{\partial }{\partial I^{i}}S_{A}\left(
\left. I\right\vert a\right) .  \label{th.p.def}
\end{equation}%
Notice that $\beta _{i}^{A}$ becomes equivalent to the standard
definition when a physical observable obeying an additive constrain:

\begin{equation}
-\frac{\partial }{\partial I^{i}}S_{A}\left( I_{T}-I\right) \equiv \frac{%
\partial }{\partial I_{A}^{i}}S_{A}\left( I_{A}\right) \equiv \beta _{i}^{A}.
\end{equation}

As already referred, equilibrium conditions (\ref{cond.eq}) allow
the practical determination of the thermodynamic parameters of the system $%
\beta _{i}^{S}$ through the corresponding parameters of the measure
apparatus $\beta _{i}^{A}$, whose interdependence with certain \textit{%
thermometric quantities} (pressure, force, length, electric signals,
etc.) with a direct mechanical interpretation is previously known.
Our next step in the present discussion is to analyze the limits of
precision of such an experimental procedure.

\subsection{Derivation of uncertainties relations}

Let us introduce the thermodynamic quantity $\eta _{i}\left(
I\right) $ defined by:
\begin{equation}
\eta _{i}\left( I\right) =-k_{B}\frac{\partial \log p_{a}\left( I\right) }{%
\partial I^{i}}=\beta _{i}^{A}-\beta _{i}^{S},  \label{def}
\end{equation}%
which is just the difference between the \textit{i}-th thermodynamic
parameters of the measure apparatus $\beta _{i}^{A}$ and the system
$\beta _{i}^{S}$ respectively. It is easy to show the validity of
the following expectation values:
\begin{equation}
\left\langle \eta _{i}\right\rangle =\int_{\Gamma }\eta _{i}\left(
I\right) p_{\alpha }\left( I\right) dI=0,  \label{exp1}
\end{equation}%
\begin{equation}
\left\langle I^{j}\eta _{i}\right\rangle =\int_{\Gamma }I^{j}\eta
_{i}\left( I\right) p_{\alpha }\left( I\right) dI=k_{B}\delta
_{i}^{j}.  \label{exp2}
\end{equation}

Derivation of these latter results reads as follows. Let us consider a set $%
\left\{ a^{i}\right\} $ of independent constants and conform with
them the
expectation value $\left\langle a^{i}A\eta _{i}\right\rangle $, being $%
A=A\left( I\right) $ certain function on the physical observables
$I$ of the
system under consideration:%
\begin{equation}
\left\langle a^{i}A\eta _{i}\right\rangle \equiv -k_{B}\int_{\Gamma
}A\left( I\right) a^{i}\frac{\partial }{\partial I^{i}}p_{\alpha
}\left( I\right) dI.
\end{equation}%
It is easy to see that the term $a^{i}\partial _{i}p_{\alpha }$ can
be rephrased as a divergence of a vector, $a^{i}\partial
_{i}p_{\alpha }= \overrightarrow{\nabla} \cdot \left(
p_{a}\overrightarrow{a}\right) $, which allows to rewrite the
above expression as follows:%
\begin{equation}
-k_{B}\int_{\partial \Gamma _{a}}Ap_{a}\overrightarrow{a}\cdot d%
\overrightarrow{\sigma }+k_{B}\int_{\Gamma
_{a}}p_{a}\overrightarrow{a}\cdot \overrightarrow{\nabla}A dI
\end{equation}%
by considering the identity $A \overrightarrow{\nabla}\cdot \left(
p_{a}\overrightarrow{a}\right) =\overrightarrow{\nabla}\cdot \left(
Ap_{a}\overrightarrow{a}\right) -p_{a}\overrightarrow{a}\cdot
\overrightarrow{\nabla}A$. The surface integral vanishes since
$p_{a}$ vanishes on the boundary $\partial \Gamma _{a}$ when $\Gamma
_{a}$ is a finite subset, or the function $A\left( I\right) $ obeys
the condition $\left\vert A\left(
I\right) \right\vert \leq C_{0}\left\vert I\right\vert ^{2}$ when $%
\left\vert I\right\vert \rightarrow \infty $, according to the
condition
(C3). Thus, we arrive at the following identity:%
\begin{equation}
\left\langle Aa^{i}\eta _{i}\right\rangle \equiv k_{B}\left\langle
a^{i}\partial _{i}A\right\rangle .
\end{equation}%
The latter result drops to Eqs.(\ref{exp1}) and (\ref{exp2}) by
considering the independent character of the set of constants
$\left\{ a^{i}\right\} $ and by assuming $A\equiv 1$ and $A\left(
I\right) \equiv I^{j}$ respectively.

Eq.(\ref{exp1}) is just the usual equilibrium conditions
(\ref{eq.cond}),
which is expressed now in terms of \textit{expectation values}:%
\begin{equation}
\left\langle \beta _{i}^{A}\right\rangle =\left\langle \beta
_{i}^{S}\right\rangle ,
\end{equation}%
or equivalently:%
\begin{equation}
\left\langle \frac{1}{T_{S}}\right\rangle =\left\langle \frac{1}{T_{A}}%
\right\rangle ,~\left\langle \left( \frac{Y}{T}\right)
_{S}\right\rangle =\left\langle \left( \frac{Y}{T}\right)
_{A}\right\rangle \label{stat.eq}.
\end{equation}%
It worth to mention that the above interpretation of thermodynamic
equilibrium conditions (\ref{eq.cond}) is more general in a
thermo-statistical viewpoint than the usual result (\ref{cond.eq})
derived from the argument of the most likely macrostate. In fact,
Eq.(\ref{stat.eq}) drops to Eq.(\ref{cond.eq}) for large
thermodynamical systems, where thermal fluctuations can be
disregarded\footnote{In absence of \textit{phase coexistence
phenomenon}, where there exist several local maxima of the
distribution function $p_{a}$.}.

Eq.(\ref{exp2}) can be rephrased as the following fluctuation relation:%
\begin{equation}
\left\langle \delta I^{j}\delta \eta _{i}\right\rangle =k_{B}\delta
_{i}^{j} \label{fluct.theo}
\end{equation}%
by considering Eq.(\ref{exp1}) and the identity $\left\langle \delta
A\delta B\right\rangle \equiv \left\langle AB\right\rangle
-\left\langle A\right\rangle \left\langle B\right\rangle $.
Uncertainty relations follows from Eq.(\ref{fluct.theo}) and the
application of the \textit{Schwarz
inequality}:%
\begin{equation}
\left\langle \delta A^{2}\right\rangle \left\langle \delta
B^{2}\right\rangle \geq \left\langle \delta A\delta B\right\rangle
^{2}
\end{equation}%
for \textit{conjugated thermodynamic quantities} $A=I^{i}$ and $B=\eta _{i}$%
. By denoting $\Delta a=\sqrt{\left\langle \delta a^{2}\right\rangle
}$, we finally obtain the following result:
\begin{equation}
\Delta I^{i}\Delta \eta _{i}\geq k_{B}.  \label{unc}
\end{equation}

\section{Direct consequences and analogies with Quantum Mechanics}

We have shown that the energy-temperature uncertainty relation:

\begin{equation}
\Delta U\Delta \left( 1/T_{A}-1/T_{S}\right) \geq k_{B},  \label{UT}
\end{equation}%
can be also extended to other conjugated thermodynamic quantities:
\begin{equation}
\Delta X\Delta \left( Y_{A}/T_{A}-Y_{S}/T_{S}\right) \geq k_{B}.
\label{XY}
\end{equation}%
A simple example is the complementarity between the internal
pressure $p_{S}$ and volume $V$ of a fluid system:
\begin{equation}
\Delta V\Delta \left( p_{A}/T_{A}-p_{S}/T_{S}\right) \geq k_{B}.
\end{equation}%
It is evident that it is impossible to determine the internal pressure $%
p_{S} $\ of a fluid system with the help of a measure apparatus
(barometer) without involving certain perturbation of its volume
$V$. Any attempt to reduce this perturbation leads to an increasing
the thermodynamic fluctuations of the quantity $\xi
=p_{A}/T_{A}-p_{S}/T_{S}$, which affects the determination of the
system pressure $p_{S}$ by using its equalization with an external
pressure $p_{A}$ associated to the measure apparatus (condition of
mechanical equilibrium).

It is important to remark that the thermodynamic uncertainty relations (\ref%
{UT}) and (\ref{XY}) are quite subtle. By considering definitions of Eq.(\ref%
{micro}), it is clearly evident that $\Delta \left( 1/T_{S}\right)
\rightarrow 0$ and $\Delta \left( Y_{S}/T_{S}\right) \rightarrow 0$ when $%
\Delta U\rightarrow 0$ and $\Delta X\rightarrow 0$. However the
corresponding thermodynamic quantities of measure apparatus $T_{A}$ and $%
Y_{A}$ \textit{become independent on the state of the system}
(defined by precise values of $U$ and $X$) at this limit situation
(thermal isolation), and hence, the system thermodynamic parameters
$T_{S}$ and $Y_{S}$ cannot be estimated by using $T_{A}$ and
$Y_{A}$. On the other hand, any attempt to change the experimental
conditions in order to reduce to zero $\Delta \left(
1/T_{A}-1/T_{S}\right) \rightarrow 0$ and $\Delta \left(
Y_{A}/T_{A}-Y_{S}/T_{S}\right) \rightarrow 0$ leads to a strong
perturbation on the state of the system $\Delta U\rightarrow \infty
$ and $\Delta X\rightarrow \infty $ and its thermodynamic quantities
$\Delta \left( Y_{S}/T_{S}\right) \rightarrow \infty $ and $\Delta
\left( 1/T_{S}\right) \rightarrow \infty $, which simultaneously
provokes a strong perturbation on the thermodynamic quantities of
the measure apparatus $\Delta \left( Y_{A}/T_{A}\right) \rightarrow
\infty $ and $\Delta \left( 1/T_{A}\right) \rightarrow \infty $.
Thus, we have arrived at the following:

\textbf{Conclusion:} \textit{It is impossible to perform an exact
simultaneous determination of the conjugated thermodynamic
quantities $I^{i}$ and $\beta^{S}_{i}=\partial_{i}S\left(I\right)$
of a given system by using an experimental procedure based on the
thermodynamic equilibrium with a second system (measure apparatus).}

In practice, we have to admit the existence of small uncertainties
in the determination of physical observables $\left( U,X\right) $
and thermodynamic parameters $\left( T,Y\right) $, which can be
disregarded in the framework
of the large thermodynamic systems where $\Delta U/U\sim 1/\sqrt{N}$, $%
\Delta X/X\sim 1/\sqrt{N}$ and $\Delta \left( 1/T\right) \sim 1/\sqrt{N}$, $%
\Delta \left( Y/T\right) $ $\sim 1/\sqrt{N}$, being $N$ the number
of system constituents. However, uncertainty relations (\ref{UT})
and (\ref{XY}) clearly indicate the limited practical utility of
some thermo-statistical concepts in systems with few constituents.

One can object that the Boltzmann-Gibbs distributions (\ref{BG})
allow a thermo-statistical description of a given system without
mattering about its size, so that, it could be always possible to
attribute, as example, a \textit{definite temperature} to any system
by using this conventional description. Basically, this is the
viewpoint of Bohr\cite{bohr}, Heisenberg\cite{Heisenberg}, as well
as other investigators\cite{Mandelbrot,Kittel}. However, such a
temperature has nothing to do with the system temperature since it
is actually the temperature of the thermostat. As already discussed
in the introductory section, the system temperature has a clear and
unambiguous definition in terms of the Boltzmann's entropy
(\ref{BE}), which has a \textit{definite value} only for a closed
system in thermodynamic equilibrium (thermal isolation) and can be
attributed to systems with few constituents. The limitation
associated to the present uncertainty relations is that any
mensuration of such a temperature (and other thermodynamic
parameters) by using the thermal equilibrium with a second system
involves an uncontrollable strong perturbation on the internal
thermodynamical macrostate of a small system, which undetermine its
initial conditions.

As already evidenced, the measure apparatus plays a role during the
experimental study of the internal state of a given system in
Statistical Mechanics quite analogous to the one existing in Quantum
Mechanics. Surprisingly, this analogy with Quantum Mechanics can be
extended to the interpretation of the uncertainty relation
(\ref{unc}) in terms of \textit{noncommuting operators}. In fact,
one can easily notice that the thermodynamic quantity $\eta
_{i}=\eta _{i}\left( I\right) $ introduced in Eq.(\ref{def}) can be
associated to a \textit{differential operator} $\hat{\eta}_{i}$
defined by:
\begin{equation}
\hat{\eta}_{i}=-k_{B}\partial _{i}
\end{equation}%
due to the validity of the mathematical identity:
\begin{equation}
\hat{\eta}_{i}p_{a}\left( I\right) \equiv \eta _{i}p_{a}\left(
I\right) . \label{ope.def}
\end{equation}%
Thus, the commutator identity $\left[
\hat{I}^{i},\hat{\eta}_{j}\right] =k_{B}\delta _{j}^{i}$, where
$\hat{I}^{i}\equiv I^{i}$, could be considered as the statistical
mechanics counterpart of the quantum relations $\left[
\hat{q}^{i},\hat{p}_{j}\right] =i\delta _{j}^{i}\hslash $.

One can provide other analogies between these physical theories.
While
Quantum Mechanics is hallmarked by the \textit{Ondulatory-Corpuscular Dualism%
}, Statistical Mechanics exhibits another kind of dualism: the one
existing between physical observables with a \textit{mechanical
significance} such as energy and other physical observables, and
those quantities with a thermo-statistical significance such as
temperature and other thermodynamical parameters. Classical
Mechanics appears as an asymptotic theory of Quantum Mechanics when
$\hbar \rightarrow 0$, while Thermodynamics
appears as a suitable approximation of Statistical Mechanics when $%
k_{B}\rightarrow 0$, or equivalently, during the imposition of the
thermodynamic limit $1/N\rightarrow 0$. Generally speaking, our
analysis seems to support the Bohr's idea about the existence of
uncertainty relations in any theory with a statistical
formulation\cite{bohr}.

\section*{Epilogue}
The present approach to uncertainty relations of Statistical
Mechanics can be considered as an improvement of
Rosenfeld\cite{Rosenfeld}, Gilmore\cite{Gilmore} and
Sch\"{o}lg\cite{Scholg} works in the past, which are also based on
fluctuation theory\cite{Landau}. In fact, our primary interest in
this thermo-statistical formulation was never related with the
justification of uncertainty relations in the framework of
Statistical Mechanics, whose long history in literature was simply
ignored by us. We accidentally advertise the existence of certain
energy-temperature complementarity during an attempt to develop an
extension of fluctuation theory compatible with the existence of
macrostates with \textit{anomalous values in response
functions}\cite{VelazquezC,Vel}, particularly, the presence of
\textit{negative heat capacities} observed in the thermodynamic
description of many \textit{nonextensive
systems}\cite{moretto,Dagostino,gro na,gro1,Dauxois}.

\section*{Acknowledgments}

It is a pleasure to acknowledge partial financial support by
FONDECYT 3080003 and 1051075. L.Velazquez also thanks the partial
financial support by the project PNCB-16/2004 of the Cuban National
Programme of Basic Sciences.

\end{document}